\documentclass[twocolumn,superscriptaddress,floatfix,preprintnumbers]{revtex4}
\usepackage{graphics,amssymb,amsmath,epsfig,color}
\usepackage{graphicx}

\begin{document}

\title{Low-dose Image Recognition with Quantum Computational Electron Microscopy}
\author{Hiroshi Okamoto}
\email[]{okamoto@akita-pu.ac.jp}
\affiliation{Department of Intelligent Mechatronics, Akita Prefectural University, Yurihonjo 015-0055, Japan}
\affiliation{Quantum and Nanotechnologies Research Centre, National Research Council, Edmonton AB T6G 2M9, Canada}
\date{\today}
\begin{abstract}
We show that quantum computational imaging is advantageous in the setting of low-dose electron microscopy of beam-sensitive specimens. Two qudits placed near the electron beam enable full transfer of quantum information between the electron microscope and a quantum computer in the proposed scheme providing the specimen is a phase object. We present a quantum algorithm that identifies the correct image among $n$ candidate images, where $n$ is larger than the effective dimension of the Hilbert space of the imaging electron.
\end{abstract}
\maketitle

Quantum computational imaging \cite{quantum_computational_imaging} is a branch of quantum \emph{computationally} enhanced measurement \cite{Aharonov_qualm, QLearning_experiments}, which also shares features with classical computational imaging \cite{miao_computational_imaging, Waller_review}. Although resolution is the principal, and often the only, figure of merit in traditional imaging, one may ask how to efficiently find the answer to a given specific question in computational imaging. An example of a specific question is whether or not a particular molecule is present in a view.

Beam-sensitive specimens present a particularly apt setting for computational imaging. The specific case we consider here is electron cryomicroscopy of a frozen biological specimen, which is easily damaged by the electron beam \cite{cryoEM_textbook}. The efficiency of measurement is measured in the number of interactions with the specimen here because that governs the specimen damage. Since too many interactions destroy the specimen, the efficiency directly translates to whether or not desired information is obtainable. 

Quantum enhancement is required, especially when only a single copy of the specimen is available. The atomic structures of many biological molecules have been obtained with \emph{classical} averaging methods such as single particle analysis, provided that a large number of copies of the molecule are available \cite{single_particle_anal_review}. In contrast, tasks such as the comprehensive identification of \emph{known} species of molecules in a crowded cellular environment, perhaps in the context of electron cryotomography \cite{template_matching}, may be a suitable arena for quantum technologies.

\emph{Quantum query complexity} captures the notion of the number of interactions with the specimen. This is the number of calls a quantum computer (QC) needs to make to an ``oracle'' to solve a problem in the abstract setting of quantum information theory \cite{query_complexity}. In the present case, the specimen is the oracle. The query model has been extensively studied in the field of quantum computing because it is relevant to many quantum algorithms, and quantum advantage tends to be provable without complexity-theoretic assumptions. Query complexity is \emph{the} figure of merit of an algorithm designed for a given task in our context, rather than a proxy for more fundamental measures such as time complexity. Thus, the advantage of quantum-enhanced measurement also tends to be provable unconditionally. 

It is also worth noting the following: In the quantum computing community, certain algorithms such as Grover's algorithm have often been characterized as offering merely a ``modest'' polynomial speedup as opposed to an exponential speedup. In contrast, in the EM community people say ``every Angstrom counts'' with respect to resolution. Hence, here is an opportunity for those modest algorithms to make a significant difference. A closely related point is that quantum advantage in our setting is free from subtleties that degrade the practicality of these modest quantum algorithms in a purely computational setting  \cite{PRX_practicality_polynomial_Qadvantage}. 

An EM capable of querying the specimen in the above general sense has been considered previously \cite{q_interface, resilient_QEM, ion_trap_QCEM}. Some of these schemes may be regarded as natural generalizations of conventional quantum electron microscopy (QEM) schemes based on the interaction between imaging electrons and a qubit \cite{eeem}. More generally, the use of quantum enhancement in EM, in order to image beam-sensitive specimens, has been extensively discussed \cite{designs_QEM,Madan_review,TEM_q_limit} and also experimental results have begun to be reported \cite{Koppell_10kV,int_free_electrons}. Many, though not all, QEM proposals aim at imaging weak phase objects beyond the shot noise limit to approach the Heisenberg limit \cite{eeem,multipass,Kaminer_spin}. Note that biological specimens are usually regarded as weak phase objects in EM.

In this Letter, we present an architecture combining a universal QC and an electron microscope and explore its uses. We call this a quantum computational electron microscope (QCEM). This essentially allows us to perform any physically possible measurement. From the fundamental physics perspective, QCEM definitely is possible. In principle, one could connect an EM to a QC via suitable quantum interfaces \cite{optical_q_interface,q_interface} placed on both the illumination and detection sides of the EM. In this way, one could transfer, or teleport, a quantum state from the QC to the illuminating electron wave to the specimen and also transfer the state of the exit electron wave back to the QC. The real question, on the other hand, is whether there exists a sufficiently simple and feasible scheme to do so. We answer this question in the affirmative, under the condition that the specimen is a pure phase object. In what follows, $-e$ denotes the electron charge. Let the $z$-axis be the electron-optical axis. Let $\lambda$ be the wavelength of the imaging electrons. A \emph{diffraction plane} is any plane conjugate to the back focal plane of the objective lens. An \emph{image plane} refers to any plane conjugate to the plane where the specimen is placed. We generally do not normalize a quantum state.

\begin{figure}
\includegraphics[scale=0.3]{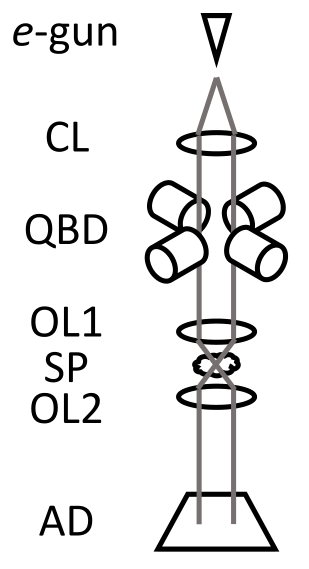}

\caption{Schematic drawing of a QCEM at the conceptual level. It comprises a pulsed electron gun ($e$-gun), condenser lens (CL), quantum beam deflector (QBD), objective pre-field lens (OL1), specimen (SP), objective post-field lens (OL2), and a pixelated area detector (AD). The electron accelerator and additional lenses, such as a projector lens, are not shown. To fully combat inelastic scattering events, energy of the scattered electron should be measured. \emph{In principle}, this can be done with time-of-flight measurement.}
\end{figure}

Figure 1 shows a QCEM scheme at the conceptual level. We defer discussions of physical realizations to later parts of this paper. At first glance, it is rather similar to the 4-dimensional (4D) scanning transmission EM (STEM) \cite{4D_STEM_review}. Following the electron gun and the condenser lens, there are two beam deflectors to bend the electron beam in the $x$ and $y$ directions at a diffraction plane. Below the pre- and post-field objective lenses and a specimen in between, a pixelated electron detector is placed at a diffraction plane. However, the crucial difference from the conventional STEM is that the beam deflectors are \emph{quantum}. Consider the beam deflector along the $x$-direction. This deflector \emph{is} a qudit, i.e., a $d$-level quantum system, with $d$ distinct quantum states $|0\rangle,|1\rangle,\cdots,|d-1\rangle$. These states are associated with, for example, a set of correspondingly equally-spaced amount of electric field, which bends the electron beam. For example, a superconducting charge qubit (the Cooper pair box) \cite{supercond_qubit_review} could be a $d=2$ version of it. The same applies to the deflector in the $y$-direction. These two qudits, which we call qudit $x$ and qudit $y$, are part of a larger QC, equipped with as many additional qubits as necessary. Entanglement-assisted QEM \cite{eeem} may be seen as the simplest version of the present scheme, with a single-axis deflector and a single-qubit QC. In EM, unlike standard quantum computation, a small-scale QC or even a single-qubit QC helps.

The effect of the quantum beam deflectors is as follows. Reflecting the $d\times d$ distinct quantum states of the combined system of qudits $x$ and $y$, there correspond $d\times d$ points on the specimen, where the electron beam is focused. Hence, one may raster-scan the electron beam by properly setting each qudit at proper times. What is newly enabled, however, is that one could also make a quantum superposition of various positions of the electron beam. Indeed, an arbitrary 2D structure of the electron beam may be generated, although it is not exactly an arbitrary structure of the electron wave front, because the electron state is heavily entangled with the two qudits. Nonetheless, it is rather remarkable that an arbitrary 2D structure can be generated by only two deflectors along the $x$ and $y$ directions. The ability of the two qudits to have entanglement between them enables this and shows that a quantum instrument could, in a sense, occasionally be simpler than the classical counterpart.

Our scheme allows for the transfer of quantum information between the probe electrons and the QC. We want to probe the specimen at $d\times d$ locations. Let these locations be indexed by two integers $p$ and $q$, where $0\le p<d$ and $0\le q<d$. Let the phase shift of the specimen at the location $\left(p,q\right)$ be $\theta_{p,q}$. For the transfer of quantum information, the availability of the following operation, which we call an ``oracle call'', is sufficient:
\begin{equation}
|p,q\rangle\Rightarrow e^{i\theta_{p,q}}|p,q\rangle,\label{eq:oracle}
\end{equation}
where $|p,q\rangle$ is a state of a quantum register in our QC, with $d^{2}$ states. (Here we somewhat enlarge the notion of an oracle from the one used in computer science, where $\theta_{p,q}$ is restricted to be either $0$ or $\pi$. Note that the latter can simulate the more widely used oracle that flips an ancilla qubit if and only if $\theta_{p,q}=\pi$.) The quantum register turns out to be the combined qudits $x$ and $y$: We let the quantum register state $|p,q\rangle=|p\rangle\otimes|q\rangle$, where $|p\rangle$ and $|q\rangle$ are the states of qudits $x$ and $y$, respectively. To realize the transform of Eq. (\ref{eq:oracle}), we first produce an electron from the electron gun in the state $|0\rangle$, so that the initial state of the combined system of the electron and the beam deflector is $|0\rangle\otimes|p,q\rangle$. Let the electron state $|n,m\rangle$ be the one that is to be focused on the location $\left(n,m\right)$ of the specimen. The action of the beam deflector is, by definition, $|0\rangle\otimes|p,q\rangle\Rightarrow|p,q\rangle\otimes|p,q\rangle$. Next, we let the electron pass through the specimen. By definition, we obtain $|p,q\rangle\otimes|p,q\rangle\Rightarrow e^{i\theta_{p,q}}|p,q\rangle\otimes|p,q\rangle$. Finally, we detect the electron in the far field. Since the electron wave from the point $\left(p,q\right)$ evolves into a plane wave in the far field, schematically we have 
\begin{equation}
|p,q\rangle=\int \frac{dk}{2\pi}\int \frac{dl}{2\pi}e^{i\left(kp+lq\right)}|k,l\rangle.\label{eq:e_diffraction}
\end{equation}
where $k$ and $l$ are real numbers that represent a point on the diffraction plane. Suppose that we detected the electron at the point $\left(k,l\right)$. This leaves the two qudits in the state
\begin{equation}
e^{i\theta_{p,q}}|p,q\rangle\otimes|p,q\rangle\Rightarrow e^{i\theta_{p,q}}\cdot e^{i\left(kp+lq\right)}|p,q\rangle.\label{eq:e_detection}
\end{equation}
Since we know $k$ and $l$ from our measurement, we can perform a phase shift operation to the qudits $x,y$ to obtain the state $e^{i\theta_{p,q}}|p,q\rangle$, which is the right-hand side of Eq. (\ref{eq:oracle}).

Next, we discuss the hardware aspect of the quantum beam deflector. We focus on qudit $x$. Consider a generalized $d$-state version of the superconducting cat qubit \cite{100_photon_qubit}. This is basically a harmonic oscillator, and we have exquisite control over its quantum state. In the QCEM context, the oscillator would most likely take the form of a microfabricated microstrip-line resonator along which the electron flies. However, here we are only interested in core physics and the resonator is modeled as a lumped LC tank circuit, and we defer numerical/geometrical details to future studies. The electron passes through the capacitor when all the energy of the harmonic oscillator is stored in the capacitor. Let the dimension of the electrodes of the parallel-plate capacitor be $l$ along the optical axis and $w$ in the perpendicular direction. Let the spacing between the two capacitor plates also be $w$ and, in keeping with the spirit of order estimation, we ignore the effect of the fringe field. The deflection angle $\theta$ must be as large as $d-1$ times $\lambda/w$, where $d$ is the number of pixels in one axis and $\lambda/w$ is the natural spread of the electron beam due to diffraction. 

We estimate the number of needed photons $N$ in the resonator to realize the deflection angle $\theta = (d-1)\lambda/w$. Let the momentum and its change of the electron be $p$ and $\Delta p$, respectively. The electron flies through the capacitor for the time duration $\tau \approx l/v = l/\beta c$, where $v$ and $c$ are the velocity of the electron and light, respectively. Let the electric field inside the capacitor be $E$. We obtain 
\begin{equation}
\theta = \frac{\Delta p}{p} = \frac{eE\tau}{p} = \frac{eEl}{p\beta c} \ge \frac{(d-1)\lambda}{w}=\frac{2\pi(d-1)\hbar}{pw}.
\end{equation}
On the other hand, $N$ satisfies $N\hbar/\sqrt{LC}=\varepsilon_0 E^2lw^2/2$ because the two expressions of energy in the capacitor must be equal. Combining these equations and assuming $C\approx\varepsilon_0 l$, we obtain 
\begin{equation}
N \ge \frac{\pi}{2\alpha}(d-1)^2\beta^2\frac{Z}{Z_0},
\end{equation}
where $\alpha\approx 1/137$ is the fine structure constant, $Z=\sqrt{L/C}$ and $Z_0 = \sqrt{\mu_0/\varepsilon_0}\approx 377\,\Omega$. We note two points here. First, $Z$ can be made significantly smaller, i.e. perhaps by an order of magnitude, than $Z_0$ in practice. Second, generation of the cat state with $\approx100$ photons in a 3-dimensional cavity has been experimentally demonstrated \cite{100_photon_qubit}. Thus, we expect that $d$ of $2\sim 3$ can be realized even for $300\,\mathrm{keV}$ electrons, and the use of slower electrons and/or the use of multiple resonators would enable a larger $d$.

Having discussed hardware designs, we proceed to consider software. In principle, any measurement method physically possible should be implementable, because of the universality of our scheme. In particular, multipass TEM \cite{multipass} is realized simply by repeated applications of Eq. (\ref{eq:oracle}), followed by phase-contrast imaging steps involving a quantum Fourier transform, its inverse and a ``phase plate'' operation in between. Note that this is a nontrivial addition to the known suite of in-focus phase contrast imaging methods in electron microscopy \cite{phase_plate_review}. A slightly better method is to apply the in-focus phase contrast imaging steps after each oracle call, which is reminiscent of Grover's algorithm. 

QCEM is strictly more powerful than conventional QEM. This point is obvious with a contrived example: Consider a phase object comprising $d^2$ pixels. Only one pixel, call it A, has a phase shift $\pi$ while all the others have zero phase shift. A straightforward application of Grover's algorithm enables identification of pixel A with $d$ passages of electrons, which clearly is impossible with conventional QEM. A subtle issue here is that pixel A receives a large amount of electron dose towards the end of Grover's search, and the procedure does not make sense if we want to minimize the localized radiation damage on pixel A. On the other hand, the method does make sense if the damage is delocalized over the entire $d^2$ pixels regardless of where the electron beam passes the specimen. (Secondary electrons often cause significant damage.) Note that if the allowed electron dose $n_d$ satisfies $d^2 \gg n_d \gg d$, QCEM is the only way to identify pixel A. 

We review several mathematical facts before proceeding. First, in a high-dimensional Hilbert space with dimension $\hat{n}$, two random unit vectors are highly orthogonal to each other. One can generate a random vector first by random walking in small steps from the origin of the space and then normalize the length of the vector. For instance, we take the real part of the $j$-th component of an unnormalized vector as a random variable $\mathbf{R}_j$ that obeys the normal distribution. The normalized unit vector has the corresponding component 
\begin{equation}
\frac{\mathbf{R}_j}{\sum_{i=1}^{\hat{n}}(\mathbf{R}_i^2 + \mathbf{I}_i^2)},
\end{equation}
where $\mathbf{I}_j$ denotes the imaginary part. Since the denominator is the sum of independent and identically distributed random variables, in the large $\hat{n}$ limit this is a constant with vanishing relative fluctuation $\approx 1/\sqrt{\hat{n}}$. Treating the denominator as a constant, we see that the real and imaginary parts of each entry of the normalized vector obey the normal distribution with the spread of $\approx 1/\sqrt{\hat{n}}$. Since each entry can be randomly positive or negative, when we take the inner product of two such random vectors, the summation amounts to a random walk for $\hat{n}$ steps with the step size $\approx 1/\hat{n}$, resulting in $\approx 1/\sqrt{\hat{n}}$ in both real and imaginary components. Second, consider the problem of deforming these randomly-generated vectors that are almost orthogonal to each other into a set of exactly orthogonal unit vectors, and how much we have to deform them. The randomly-generated $\hat{n}$ unit vectors $|v_1\rangle, |v_2\rangle, \cdots,|v_{\hat{n}}\rangle$ are almost orthogonal to each other in the sense that $|\langle v_i|v_j\rangle|<\delta$ for a small $\delta$. We deform these to obtain orthogonal unit vectors $|u_1\rangle, |u_2\rangle, \cdots,|u_{\hat{n}}\rangle$. Let the standard basis vectors be $|1\rangle, |2\rangle, \cdots,|\hat{n}\rangle$. Define $u_{ij} = \langle i|u_j\rangle$ and $v_{ij} = \langle i|v_j\rangle$ and let $U,V$ be square matrices with the components $u_{ij}$ and $v_{ij}$, respectively. Due to the orthogonality of $|u_i\rangle$ vectors, we have $U^\dagger U = I$, where $I$ is the unit matrix. Similarly, due to the approximate orthogonality, we have the Gram matrix $G = V^\dagger V = I + E$, where the absolute values of the off-diagonal elements of $E$ are smaller than $\delta$ while the diagonal elements are zero. Verify that the deformation $U = VG^{-1/2}$ gives $U^\dagger U = I$. An important question is how close are the set of vectors $|v_i\rangle$ and $|u_i\rangle$. To this end, we find $\langle u_i|v_i\rangle$, which are the diagonal elements of 
\begin{equation}
U^\dagger V = G^{-1/2}V^\dagger V = (I + E)^{1/2} \approx I + \frac{E}{2} - \frac{E^2}{8} +\cdots.
\end{equation}
From this expression, we find that the diagonal element is $1 - (\hat{n} - 1)\delta^2/8 + \cdots$.

Consider the practically relevant problem of identifying the correct image among $n$ candidate images. Here, candidate images are represented as quantum states that are the result of the multipass process with the candidate specimen structures. First, consider the simple case of $n = d^2$. We assume that all candidate states are almost orthogonal to each other as random vectors are. If not, we deform the nearly-orthogonal set of candidate quantum states to make it a strictly orthogonal set. In this case, the dimension of the Hilbert space for the electron quantum state equals the number of candidates, and we can simply measure the quantum state once with respect to the measurement basis set comprising the candidate states. 

The number of candidate images $n = d^2$ is expected to be too small for practical purposes. There are far too many possible images, even if we restrict the search space to ``natural'' images amidst the vast array of white noise images. For example, the same structure with an arbitrary amount of spatial displacement in the image plane is enough to generate $d^2$ candidate images. On the other hand, Holevo's bound dictates that we cannot increase the number of candidates for a measurement with a single quantum state living in the $d^2$-dimensional Hilbert space. Hence, we must use $m > 1$ copies of the same quantum state.

We begin with a less-than-ideal measurement strategy. Consider $m > 1$ sets of measurement basis states, which we call the measurement bases hereafter, in the $d^2$-dimensional Hilbert space. Within each measurement basis, by definition all the basis states are strictly orthogonal to each other. Two vectors from different measurement bases are expected to be almost orthogonal to each other in high dimension, but they cannot be strictly orthogonal because one cannot fit more than $d^2$ orthogonal vectors in the $d^2$-dimensional Hilbert space. The correct image may belong to one particular, ``relevant'' measurement basis. Our first measurement strategy for the case $n > d^2$ may be to measure the state with one measurement basis after another, and repeat the entire process. If a particular measurement basis produces the same measurement outcome repeatedly, this measurement basis is expected to be ``relevant'' and the measurement outcome is likely to correspond to the correct image. We expect false positives when $n > d^4$. 

A better strategy is to perform a \emph{coherent} measurement over all $m$ quantum states obtained in the same way \cite{QLearning_experiments}. In other words, the final measurement is performed on the entire Hilbert space, which is the tensor product of all $m$ constituent Hilbert spaces. Exponentially many, i.e. $d^{2m}$, basis states can live in this Hilbert space, although our states are restricted to the $m$-fold tensor product of a single, identical state. The overlap between formerly almost-orthogonal vectors $|\langle w_i|w_j\rangle| < \varepsilon$ gets exponentially suppressed as 
\begin{equation}
\left\{\bigotimes_{k=1}^m\langle w_i|\right\}\left\{\bigotimes_{k=1}^m|w_j\rangle\right\}< \varepsilon^m.
\end{equation}
Hence, we may deform them by 
only an exponentially small amount to obtain a set of basis vectors that are exactly orthogonal to each other. With this expanded set of measurement basis states, we expect to obtain the correct image in one shot. Note that this measurement has certain robustness to inelastic scattering processes that often destroy quantum measurement: Since we obtain $m$ quantum states independently from each other, if one quantum state acquisition fails, we may simply discard the corrupt quantum state and try again. 

Many thanks to Professors Marek Malac and Robert M. Glaeser for useful discussions. A significant part of this research was done while the author was on sabbatical in Edmonton. The author thanks both the accepting and home institutes for their support. Chatbots Claude and Gemini have been used for brainstorming. This research was supported in part by the JSPS ``Kakenhi'' Grant (Grant Numbers 19K05285, 25K07783).

\end{document}